\documentclass[%
 reprint,
 amsmath,amssymb,
 aps,
]{revtex4-2}

\usepackage{graphicx}
\usepackage{latexsym}
\usepackage{amsmath}
\usepackage{amssymb}
\usepackage{dblfloatfix}
\usepackage{amsfonts}
\usepackage{xcolor}
\usepackage{cases}
\usepackage[normalem]{ulem}

\begin{document}

\preprint{APS/123-QED}

\title{KPZ scaling in one-dimensional arrays of photon condensates}

\author{Michiel Yzewyn}
\author{Michiel Wouters}%
\affiliation{%
 TQC, Universiteit Antwerpen, Universiteitsplein 1, B-2610 Antwerpen, Belgium
}%

\date{\today}

\begin{abstract}

Bose-Einstein condensates of photons in dye filled microcavities are known to feature many properties of the ideal bose gas at thermal equilibrium. Nevertheless, the finite photon life time renders them driven dissipative systems where losses are balanced by continuous driving by a pump laser. We show here with simulations based on a classical field model that their driven-dissipative nature causes their first order coherence to feature Kardar-Parisi-Zhang (KPZ) scaling, a phenomenon well known in the related condensates of exciton-polaritons. Remarkably, we find that the KPZ scaling even occurs in a regime with large density fluctuations. We discuss the modification of the scaling by the occupation of excited states and how to recover the universal scaling behavior by spectral filtering.

\end{abstract}

\maketitle


\section{Introduction}

Universality is a powerful concept that allows one to identify profound similarities between microscopically disparate systems. Originally introduced for equilibrium systems close to second order phase transitions, it also applies to scaling in systems out of equilibrium. In the latter context, the universality class of the Kardar-Parisi-Zhang (KPZ) equation \cite{kardar1986dynamic,halpin1995kinetic,halpin2015kpz} has turned out to play an important role. First proposed to describe the dynamics of stochastic interface growth \cite{kardar1986dynamic}, with a nonlinear term expressing their propensity to grow in the direction that is locally perpendicular to the surface, it has proven to be a fertile field of mathematical study and is experimentally relevant for systems ranging from liquid crystal interfaces \cite{Takeuchi_2012} to firefronts and cell growth \cite{CellGrowthKPZ}.

By mapping the phase of a quantum fluid to the surface height, the KPZ equation also appears to govern the long wavelength and low frequency phase fluctuations of out-of-equilibrium quantum fluids \cite{gladilin2014spatial,ji2015temporal,altman2015two,sieberer2016lattice,zamora2017tuning,squizzato2018kardar,fontaine2022kardar}. These fluctuations are experimentally reflected in the long-time and large-distance decay of the spatiotemporal coherence \cite{fontaine2022kardar,widmann2025observation}, that feature a scaling collapse onto the universal KPZ scaling function \cite{prahofer2004exact}.

Experimentally, non-equilibrium quantum fluids are realized in photonic platforms, such as exciton-polariton condensates in semiconductor microcavities \cite{Kasprzak:Nature2006} and photon condensates in dye filled cavities \cite{klaers2010bose,marelic2015experimental,greveling2018density} or semiconductors \cite{pieczarka2024bose,barland2021photon,schofield2024bose}. All these systems are essentially variations on spatially extended lasers \cite{bloch2022non}, that differ in their microscopic mechanisms, but share the same order parameter and U(1) symmetry preserving drive and dissipation. In exciton-polariton systems, the gain that compensates for the finite polariton life time comes from exciton scattering into the lower polariton region, whereas for photon condensates, the gain stems from the emission of excited dye molecules or electron-hole recombination.
Both platforms have witnessed impressive technological advances over the last decades, resulting in photonic cavities with negligible disorder that are pumped in a versatile way and whose coherence properties can be inferred with great precision.

For what concerns measurements of the spatiotemporal coherence, the exciton-polariton platform has taken the lead \cite{fontaine2022kardar}, but there is at first sight no fundamental reason why photon condensates should not be suitable for the study of KPZ scaling. One of the latter platform's advantages is that the microscopic dynamics governed by the emission and absorption spectra of well characterized dyes is very well known, in contrast to the more complex relaxation processes that are involved in exciton polariton condensation. Therefore, photon condensates can be modeled theoretically at a quantitative level with very few fitting parameters \cite{klaers2010bose,sazhin2024observation}.

One aspect that constitutes a marked difference between polariton and photon condensates is the magnitude of density fluctuations. Polariton condensates follow the usual laser like behavior with second-order coherence function at zero delay $g^{(2)}(\tau=0)=2$ below and $g^{(2)}(0)=1$ above threshold \cite{adiyatullin2015temporally,kim2016coherent}. In photon condensates on the other hand, the good chemical equilibrium between photons and excited molecules, together with the absence of photon-photon interactions, imply that the density fluctuations can approach the large grand canonical density fluctuations of the ideal bose gas (IBG) featuring $g^{(2)} = 2$, also above the BEC threshold \cite{schmitt2014observation}. It was pointed out in Ref. \cite{loirette2023photon} that the large density fluctuations are not due to the thermalization itself, but are rather to be attributed to a weak gain saturation nonlinearity.


Theoretical descriptions of photon condensates have been developed at various levels of complexity. On the one hand, descriptions based on the quantum master equation \cite{kirton-keeling-13} are the most rigorous but at the same time the most computationally demanding. Due to the prohibitively large size of the Hilbert space for spatially extended systems, the full quantum description can not be used without further (drastic) approximations. 
Semiclassical Boltzmann rate equations on the other hand provide the most basic theoretical framework \cite{schmitt2018,walker2019collective,loirette2023photon}.
This approach, however, does not describe the phase fluctuations and is therefore unable to capture KPZ scaling. A middle ground is formed by a stochastic classical field description for the photons, coupled to rate equations for the excited molecules \cite{gladilin2020classical}, in analogy with the truncated Wigner descriptions of polariton condensates \cite{Wouters:PRB2009}. The supercurrents due to phase gradients are described by the Hamiltonian terms, where the dissipative terms represent gain and loss. 

In this article, we present a theoretical study of KPZ scaling in one dimensional photonic condensates based on simulations with the classical field model. In order to avoid issues related to the UV cutoff, we consider a lattice geometry, which is experimentally realized by modulating the mirror thickness or the refractive index \cite{dung2017variable,vretenar2021controllable}.

As expected on the grounds of universality, we do find KPZ behavior in the scaling of the spatiotemporal coherence. Strikingly, the numerical results show that KPZ scaling emerges not only in the regime of small density fluctuations, but also in the regime of large density fluctuations. In both cases the spatial and temporal scaling of the correlation functions are consistent with KPZ predictions. However, we find that a collapse of spatiotemporal coherences onto the universal KPZ scaling curve is obtained only after filtering out the high-energy modes. 

The article is organized as follows. In Sec. \ref{sec:model} the stochastic classical field model is discussed from which the KPZ equation is derived analytically in Sec. \ref{sec:analytical}. Numerical results obtained from the model are shown in \ref{sec:numerics} for both large density fluctuations (grand canonical regime) and medium density fluctuations (crossover canonical regime). Our conclusions and outlook are given in \ref{sec:conclusion}.

\section{Model}\label{sec:model}

A classical stochastic description of the site and time dependent photon amplitude $\psi(x,t)$ in a one dimensional array of coupled cavities is  (in units with $\hbar =1$) given by \cite{gladilin2020classical}
\begin{align}
i \frac{\partial \psi(x,t) }{\partial t}  = 
&- (1-i\kappa) J  \left[ \psi(x-1,t) +\psi(x+1,t) - 2 \psi(x,t) \right] \nonumber \\
&+
\frac{i}{2}\left[ B_{21}M_2(x,t) - B_{12}
M_1(x,t)-\gamma\right]\psi(x,t) \nonumber \\
&+ \sqrt{D(x,t)} \; \xi(x,t).
 \label{eq:gGP}
\end{align}
Here, $J$ is the tunneling rate between neighboring cavities, $\gamma$ is the cavity loss rate and $B_{12}\, (B_{21})$ is the Einstein absorption (emission) coefficient of the dye at the frequency of a single cavity. The number of ground state and excited dye molecules is denoted by $M_1$ and $M_2$ respectively.
The Kennard-Stepanov relation \cite{kennard,stepanov,moroshkin} between emission and absorption results in the term that is proportional to the dimensionless parameter $\kappa$,  equal to
\begin{equation}
    \kappa = \frac{B_{12} M_1 }{2 k_B T},
\end{equation}
with $T$ the temperature and $k_B$ the Boltzmann constant. The terms on the r.h.s. of Eq. \eqref{eq:gGP} that are proportional to $\kappa$ correspond to the model A relaxational dynamics that drives the system into the thermal equilibrium state \cite{hohenberg1977theory}, but because the real and imaginary parts of the coefficients on the r.h.s. of Eq. \eqref{eq:gGP} do not coincide, the steady state is fundamentally a nonequilibrium one \cite{sieberer_2013}. 

The last term in Eq. \eqref{eq:gGP} represents complex Gaussian noise, where $\xi$ has zero mean and the autocorrelation function $\langle \xi^*(x,t) \xi(x',t') \rangle  =  \delta_{x,x'} \delta(t-t')$. The noise originates from the dissipative processes and has magnitude $D(x,t) = (B_{12} M_1(x,t) + B_{21} M_2(x,t) + \gamma)/2$. Because under experimental conditions for photon condensation, the relative population inversion $(M_2-M_1)/M$, with $M$ the total number of molecules, is much smaller than one, we replace the time-dependent noise with the time-independent one $D=B_{12} M+\gamma/2$, making the noise term in Eq. \eqref{eq:gGP} of additive nature, which is numerically easier to work with.
In the absence of losses, the noise strength satisfies the fluctuation-dissipation relation with the friction parameter $\kappa$.

The local change in photon number $n(x,t) = |\psi(x,t)|^2$, except for the one caused by the cavity losses, originates from molecular transitions between the ground and excited states. 
The photon field equation is therefore coupled to a rate equation for the number of excited molecules
\begin{align}
\frac{\partial }{\partial t} M_2(x,t) &= - \frac{\partial}{\partial t} |\psi(x,t)|^2 - 2J \Im[\psi^*(x,t)\frac{\partial}{\partial x}\psi(x,t)] \nonumber \\ &- \gamma  |\psi(x,t)|^2 + P ,
\label{eq:rate}
\end{align}
where the last term represents the incoherent pumping of the molecules that replenishes the excitations that are lost through photon losses. From the balance between pump and loss, one obtains the relation $P=\gamma \bar n$, where $\bar n$ is the average photon number.

Some first insight into the fluctuation properties of photon condensates can be obtained by first neglecting the spatial degrees of freedom (single mode approximation) and considering the photon number fluctuations $\delta n=n-\bar n$ in the absence of losses and in the linear approximation. The stochastic equation of motion then reads
\begin{equation}
\frac{d}{dt} \delta n = -\Gamma\, \delta n + \sqrt{D_n}\, \xi_n,
\label{eq:deltanhom}
\end{equation}
where the real Gaussian noise $\xi_n$ has the autocorrelation function $\langle \xi_n(x,t) \xi_n(x',t') \rangle  =  \delta_{x,x'} \delta(t-t')$ and magnitude $D_n = 2 B_{12} M \bar n$.

The number fluctuation decay rate $\Gamma$ is given by 
\begin{equation}
\Gamma = \left(1+\frac{n^2}{M_{\rm eff}}\right)\frac{B_{12} M}{\bar n},
\label{eq:Gamma}
\end{equation}
where, the effective reservoir size equals
\begin{equation}
M_{\rm eff} = \frac{ M}{2[1+\cosh(\beta\Delta)]}.
\end{equation}

Fom Eq. \eqref{eq:deltanhom}, we obtain for the density fluctuations \cite{verstraelen2019temporal}
\begin{equation}
\langle \delta n^2 \rangle = \frac{\bar n^2}{1+\frac{\bar n^2}{M_{\rm eff}}}.
\label{eq:densfluct}
\end{equation}
This allows to distinguish between two regimes \cite{schmitt2014observation}, the grandcanonical with large density fluctuations when $M_{\rm eff} \gg \bar n^2$ and the canonical with small density fluctuations for  $M_{\rm eff} \ll \bar n^2$.

In a bose gas with significant density fluctuations, a conceptually important quantity is the \textit{quasi-}condensate density, defined as \cite{prokof2001critical} $n_{\rm qc} = \sqrt{\bar n^2 - \delta n^2}$. 
In the grandcanonical regime, the quasi-condensate density is always vanishingly small. In the canonical regime, there is a large quasi-condensate density in the single mode case, which may be reduced in the multimode case due to photons hopping between the cavities.

In terms of the quasi condensate density, the photon field can then be written as \cite{svistunov2015superfluid}
\begin{equation}
    \psi = \sqrt{n_{\rm qc}}\,e^{i\theta} + \delta \psi.
    \label{eq:quasicond}
\end{equation}
Here $\delta \psi$ is a Gaussian field that represents the short range fluctuations \cite{svistunov2015superfluid}. The long distance and/or time behavior of the correlations is governed by the phase fluctuations:
\begin{equation}
   \langle \psi^*(x,t) \psi(0,0) \rangle\; \xrightarrow{{\rm large} \; x, t} \;  n_{\rm qc}\, \langle e^{i[\theta(0,0)-\theta(x,t)]} \rangle.
\end{equation}
In the second cumulant approximation, the contribution of the phase fluctuations to the decay of the spatio-temporal correlations equals
\begin{equation}
     \langle \psi^*(x,t) \psi(0,0) \rangle\ \approx \, n_{\rm qc}\,
     e^{- \frac{1}{2} \left\langle \left[ \theta(x,t)-\theta(0,0) \right]^2\right\rangle  },
    \label{eq:correlator}
\end{equation}
showing that the roughness of the phase directly translates to the coherence decay.

\section{Analytical derivation of KPZ phase dynamics \label{sec:analytical}}

The coupled set of equations \eqref{eq:gGP} and \eqref{eq:rate} are analogous to the ones that are used to describe polariton condensates. For the inchoherently pumped polariton system, the KPZ equation has been theoretically derived \cite{gladilin2014spatial,ji2015temporal,altman2015two,sieberer2016lattice,zamora2017tuning,squizzato2018kardar,fontaine2022kardar} and the resulting spatiotemporal scaling was experimentally observed in both one and two dimensional systems \cite{fontaine2022kardar,widmann2025observation}. 

Based on the close analogy, one therefore expects that also the driven dissipative photon condensates fall into the KPZ universality class. A first argument in favor of this expectation can be constructed by considering the density and phase fluctuations on top of a uniform condensate: $\psi(x,t) = \sqrt{\bar n +\delta n(x,t) } \, e^{i \theta(x,t)}.$ 

The KPZ dynamics of the phase can then be derived by eliminating the density fluctuations under the assumption that they are small, an assumption that is not valid in all regimes of photon condensation but that we will make here anyway. Because the KPZ scaling should occur at large distances and times, we go for simplicity to the continuum version of the model \eqref{eq:gGP} and make the replacement $ J  \left[ \psi(x-1,t) +\psi(x+1,t) - 2 \psi(x,t)  \right] \rightarrow -\nabla^2 \psi(x,t)/2m$, with the band mass $m=1/2J$, where we have taken the lattice spacing as our unit of length.

The linearized equation of motion for the deviations in the photon density and total excitation number $X = n+M_2$ reads
\begin{align}
\frac{\partial}{\partial t} \delta n &=   -\frac{\bar n }{m}\nabla^2 \theta  - \left(\Gamma - \frac{\kappa }{2m }   \nabla^2 \right) \delta n + \alpha \Gamma \delta X \label{eq:deltan}\\
\frac{\partial}{\partial t} \delta X &=  -\frac{\bar n }{m}\nabla^2 \theta-\gamma \delta n
\label{eq:deltaX}
\end{align}
where $\alpha=\partial \bar n/\partial X$.

Under the assumption that the photon and reservoir density fluctuations relax much faster than the phase fluctuations, we can set the time derivatives in Eqs. \eqref{eq:deltan} and \eqref{eq:deltaX} equal to zero and obtain from Eq. \eqref{eq:deltaX}
\begin{equation}
\delta n \approx  - \frac{\bar n}{m \gamma } \nabla^2 \theta.
\end{equation}
Substituting these density fluctuations into the equation of motion for the phase 
\begin{align}
    \frac{\partial \theta}{\partial t} &= 
    \frac{1}{2m}\left[-\frac{(\nabla \delta n)^2}{4 n^2}+\frac{\nabla^2 \delta n}{2n}-(\nabla \theta)^2
    \right] \nonumber \\ 
    &+ \frac{\kappa}{2m}\left[ 
    \frac{\nabla \delta n \nabla \theta}{n}+\nabla^2 \theta
    \right]
    + \sqrt{D_\theta} \xi_\theta,
    \label{eq:phase0}
\end{align}
and retaining only the terms that dominate at long wave lengths yields 
\begin{equation}
\frac{\partial}{\partial t} \theta = \frac{\kappa}{2m} \nabla^2 \theta - \frac{1}{4\gamma m^2} \nabla^4 \theta - \frac{1}{2m} (\nabla \theta)^2 + \sqrt{D_\theta}\, \xi_{\theta}.
\label{eq:KS}
\end{equation}
The real Gaussian phase noise is of magnitude $D_\theta = D/\bar n$ and $\xi_\theta$ has autocorrelation $\langle \xi_\theta(x,t) \xi_\theta(x',t') \rangle  =  \delta_{x,x'} \delta(t-t')$.
At long wavelengths, the fourth order derivative can be neglected and one obtains the KPZ equation for the phase dynamics, but keeping it does not change the universality class.

Mapping the parameters of phase equation \eqref{eq:KS} onto the KPZ equation \eqref{eq:KPZ_OG} where $<\eta\:\eta'> = D\delta_{x,x'}\delta_{t,t'}$, allows for an analytical derivation of the scaling coefficients $y_0$ and $c_0$ of eq. \eqref{eq:scaling}.

\begin{align} \label{eq:KPZ_OG}
    \frac{\partial h(x,t)}{\partial t} = \nu \nabla^2h(x,t) + \frac{\lambda}{2}(\nabla h(x,t))^2 + \eta(x,t)
\end{align}

Identifying the terms in eq. \eqref{eq:KS} yields:
$\nu = \kappa/(2m)$, $\lambda=1/m$ and $D =\mathcal{D}/n=\kappa T/n$. 

When the fourth order derivative in equation \eqref{eq:KS} is neglected, the steady state phase-phase correlations are no longer affected by the nonlinear term and are given by their thermal equilibrium expression
\begin{equation}
    \langle [\theta(\Delta x)- \theta(0)]^2  \rangle = \frac{T m}{\bar n} \Delta x.
\end{equation}
Therefore, at small density fluctuations and within the second cumulant expansion, this leads to the photonic spatial coherence
\begin{equation}
    g^{(1)}(\Delta x) = \frac{1}{\bar n}\langle \psi^*(\Delta x) \psi(0)\rangle 
    \approx \langle e^{-\frac{1}{2}[\theta(0)-\theta(\Delta x)]^2} \rangle
    = e^{ -  \Delta x /\ell_c }
    \label{eq:g1_analytical}
\end{equation}
with the same coherence length $\ell_c = 2 \bar n/T m$ as for the weakly interacting  Bose gas at temperatures larger than the interaction energy  \cite{BECbook}.
Even though the interaction strength does not explicitly appear in the expression of the correlation length of the weakly interacting bose gas, it is by a factor of two larger than the correlation length of
the noninteracting Bose gas (NIBG) $\ell^{\rm ideal}_c  = \bar n /T m$, where density fluctuations are large.
As we will see in the next section, the noninteracing photon condensates have the correlation length of the NIBG, indicating that while they obey the linear scaling of the phase-phase correlator with distance,  the approximations leading to Eq. \eqref{eq:KS} are  not valid. Fortunately, the robustness of universality does not impede the spatiotemporal scaling to fall in the KPZ class.

\section{Numerical simulation results}\label{sec:numerics}

The model's exact dynamics are investigated by numerically sampling the coupled nonlinear equations \eqref{eq:gGP} and \eqref{eq:rate}.
All simulations have been carried out in the Julia programming language \cite{juliaLang} using the DifferentialEquations package \cite{diffeqjul}. 
The numerical simulations allow us to investigate the regimes where the density fluctuations are not small. As discussed above Eq.~\eqref{eq:densfluct}, the strength of the density fluctuations due to the exchange of excitations between the photons and molecular excitations can be tuned by the parameter $n^2/M_{\rm eff}$. Below we will address both the grand-canonical and canonical regimes cases consecutively.

\subsection{Grand canonical regime}
We start with a parameter set for which $n^2/M_{\rm eff}\approx 1.3 \times 10^{-4}$, which is according to Eq. \eqref{eq:densfluct} deeply in the grand canonical regime where the quasi condensate fraction is much smaller than one. Numerically, we have obtained $n_{\rm qc}\approx 0.008$, which is not too far from the analytically expected value of $0.01$.
We took the emission and absorption coefficients of the experimentally used dye \cite{klaers2010bose,schmitt2018} but a worse quality of the mirrors than usually used for the study of photon condensates in order to enhance the nonequilibrium nature of the condensate. The ratio of the reabsorption rate to the cavity life time, the Knudsen parameter \cite{loirette2023photon}, equals for our parameters $\gamma/(B_{12} M_1) \approx 10^{-5}$, indicating very good thermalization. 

\begin{figure}[h!]
    \begin{minipage}{.5\linewidth}
        \centering
        \includegraphics[width = 1.0\linewidth]{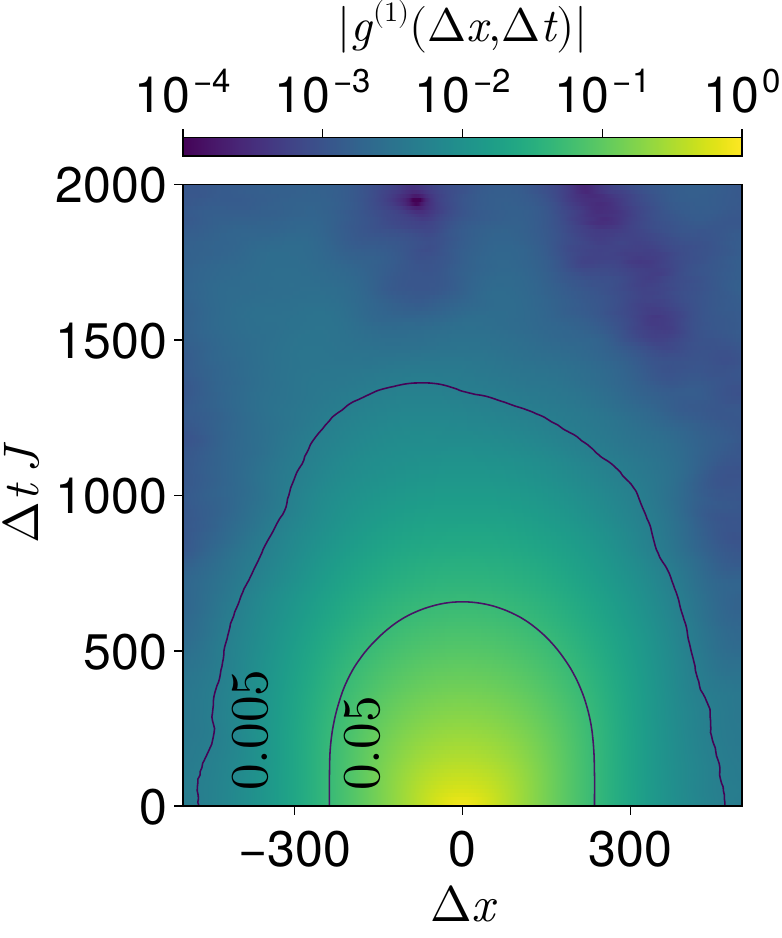}
    \end{minipage}%
    \begin{minipage}{.5\linewidth}
        \centering
        \includegraphics[width=1\linewidth]{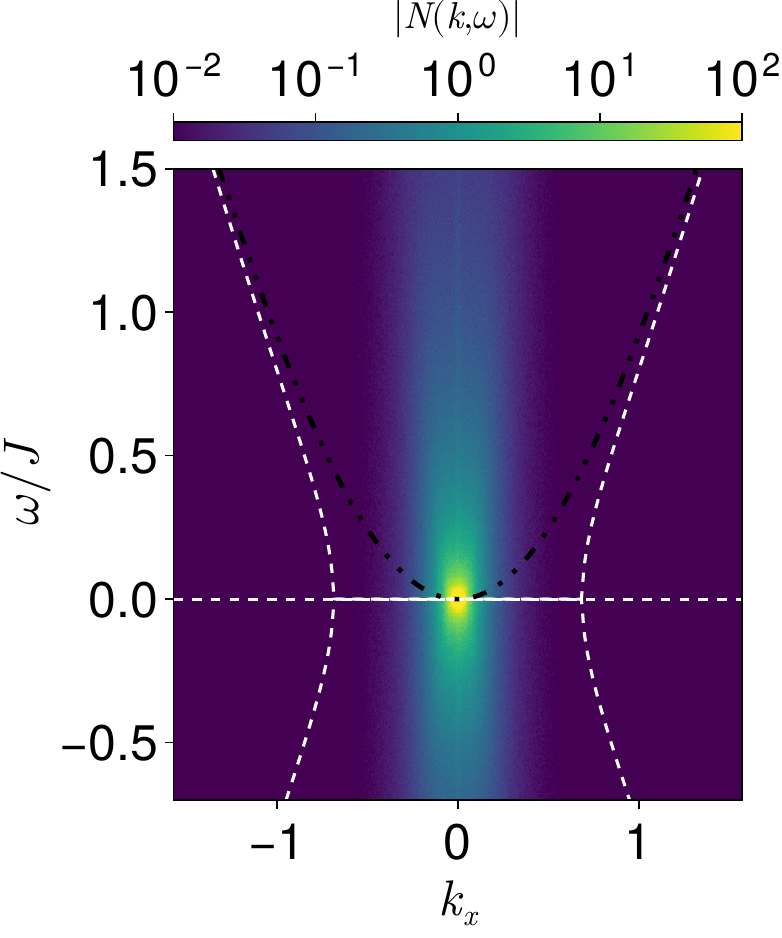}
    \end{minipage}\par\medskip
    \caption{\textit{Left panel --} Spatiotemporal first order coherence of the photon condensate in the grandcanonical regime. The area inside the contours shows the spacetime points collapsing on the universal KPZ scaling curve for the grand canonical regime with $0.005 \le |g^{(1)}(\Delta x, \Delta t)| \le 0.05$ (see Fig. \ref{fig:scaling}). \textit{Right panel} -- The energy-momentum spectrum of the photon condensate is shown in false color scale, the black dashed line represents the photon dispersion of the empty cavity, the white dashed lines the elementary excitation spectrum.
    System parameters: $J=1\:meV, \Delta=-60\:J, \gamma=0.01\:J, B_{21}=10^{-7}\:J, n=10^3, M=10^{11}$.
    }
    \label{fig:spectrumcohmapgrandcan}
\end{figure}

\begin{figure}[h!]
    \centering
    \includegraphics[width = 1.0\linewidth]{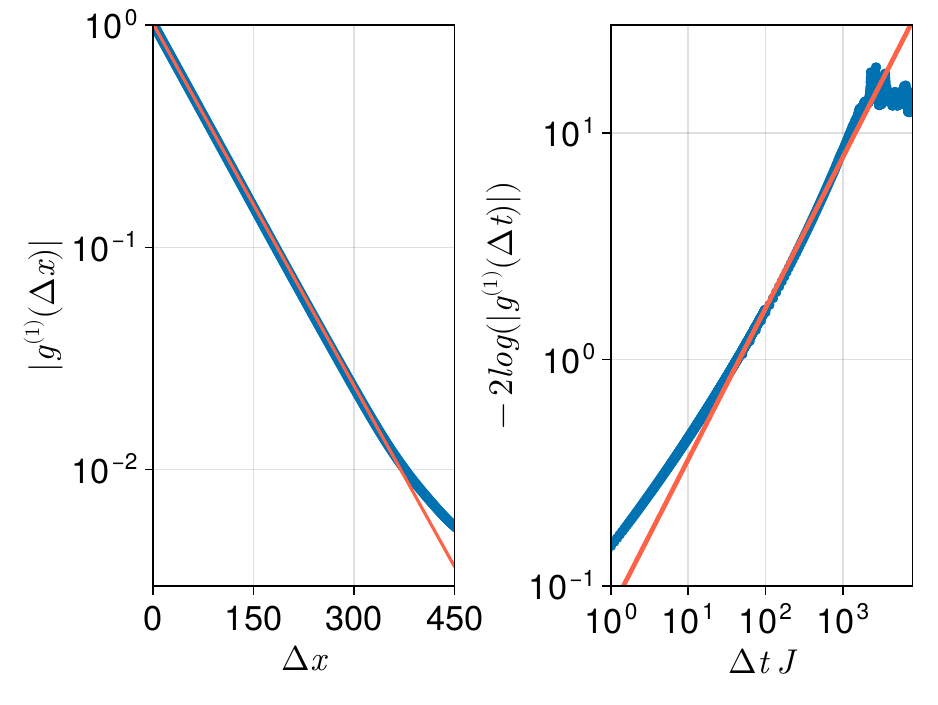}
    \caption{First order spatial (left) and temporal (right) coherence decay for a one dimensional photon condensate array in the grand canonical regime. Numeric result (blue data) from simulating eq. \eqref{eq:gGP} and \eqref{eq:rate} for equal-time spatial correlation function $g^{(1)}(x)$ (left) and equal-space temporal correlation function $g^{(1)}(t)$ (right) for a $N=1024$ site system with periodic boundaries in the grand canonical regime: $n^2/M_{\rm eff}\approx 1.3 \times 10^{-4}$. |Left: The red line depicts the NIB exponential decay $\ell_c  = \frac{2J \bar n}{T}$. The numeric data follows the NIB decay up to a distance when FFT finite size effects start to affect the simulation. |Right: The red line depicts the KPZ temporal scaling curve $a(\Delta tJ)^{2/3}$ fit to the numeric data with fitting parameter $a$. The numeric data follows KPZ scaling up to late times when the simulation is not statistically converged. Same system parameters as in Fig. \ref{fig:spectrumcohmapgrandcan}.}
    \label{fig:spatcorgrandcan}
\end{figure}

The resulting spectral density of the photons is shown in Fig. \ref{fig:spectrumcohmapgrandcan}. The spectrum is clearly dominated by the condensate peak, with no visible population of excited states.
The corresponding spatiotemporal coherence is shown in Fig. \ref{fig:spatcorgrandcan}.
It decays smoothly in both the temporal and spatial directions. 

As discussed below Eq. \eqref{eq:g1_analytical}, in the regime of large fluctuations the spatial correlation length is expected to correspond to the one of the NIBG  $\ell^{\rm ideal}_c  = \bar n /T m$, which reads in terms of the tunneling strength  $\ell^{\rm ideal}_c = 2 J \bar n /T$, where we use the lattice spacing as our unit of length.  
Fig. \ref{fig:spatcorgrandcan} compares the decay of the spatial coherence from the numerical simulations with this prediction, showing very good agreement. 
Since the coherence length equals $\ell_c=80$, we did the simulations on a large lattice of 1024 sites in order to limit the finite size effects and for the system to be large enough for the KPZ scaling to develop.
Exponential decay of spatial correlations is in one dimensional systems so ubiquitous that, while compatible with KPZ scaling, it is definitely not a particular hallmark.

The decay of the temporal correlations on the other hand, reflects the dynamical exponent $\beta$, that does discriminate between KPZ ($\beta=1/3$) and other scaling behaviors, such as Edwards-Wilkinson scaling ($\beta=1/4$) which is obtained when the nonlinear term in the KPZ equation is omitted \cite{halpin1995kinetic}. Fig. \ref{fig:spatcorgrandcan} shows the behavior of the temporal coherence. It indeed features the growth exponent $\beta=1/3$ in the temporal coherence, which is according to \eqref{eq:correlator} a direct consequence of the temporal scaling of the phase fluctuations as
\begin{equation}
   \left \langle [\theta(0,\Delta t) - \theta(0,0)]^2 \right \rangle \sim \Delta t^{2\beta} = \Delta t^{2/3}.
\end{equation}
The numerically determined temporal coherence tends to this power law at times later than the initial nonuniversal short time decay, but not too late because there our numerical results are not statistically converged. The expected late time Shawlow-Townes scaling for finite system sizes \cite{amelio2020} is not observed in our numerics because of too large statistical noise.

\begin{figure}[h!]
    \centering 
    \includegraphics[width = 1.0\linewidth]{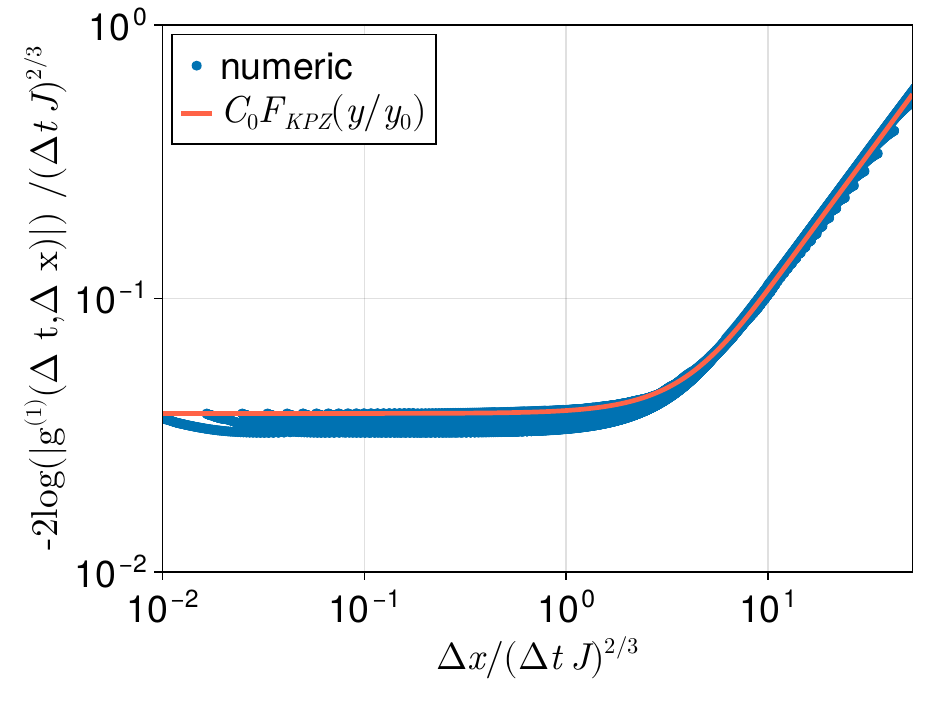}
    \caption{The rescaled numerical data points for intermediate values of the coherence, taken between the contours in the left panel of Fig. \ref{fig:spatcorgrandcan}, show a good collapse, that agrees very well with the universal KPZ scaling curve. Fitted parameters: $y_0\approx8.3\times10^{-1}, C_0\approx3.0\times10^{-2}$. Same system parameters as in Fig. \ref{fig:spectrumcohmapgrandcan}.}
    \label{fig:scaling}
\end{figure}

The full KPZ spatiotemporal scaling is expressed by the scaling function $F_{\rm KPZ}$:
\begin{equation}
    \langle \theta(\Delta x,\Delta t) \theta(0,0) \rangle = C_0 \; \Delta t^{2/3} F_{\rm KPZ}(y/y_0),
    \label{eq:scaling}
\end{equation}
with $y=\Delta x/\Delta t^{1/z}$,
that holds for all distances and times that are large enough to fall in the universal scaling regime. 
Because of the imperfect convergence of our numerical results, we also have to eliminate the points at too large distances and times. The space-time points that were retained for producing the scaling collapse are the ones that fall in Fig. \ref{fig:spectrumcohmapgrandcan} between the two contours at $g^{(1)}=0.05$ and $g^{(1)}=0.005$.
In Eq. \eqref{eq:scaling}, $C_0$ and $y_0$ are nonuniversal parameters, that we determine by fitting to the simulation data.
Figure \ref{fig:scaling} shows very good collapse of the numerically determined spatio-temporal coherence and an excellent agreement between the theoretically \cite{prahofer2004exact} determined one-dimensional KPZ scaling function, further corroborating that the phase dynamics of a photon condensate falls in the KPZ universality class, even when it has large density fluctuations.

As could be expected in the large fluctuation regime, we find a big discrepancy between the numerically fitted scaling coefficients $(C_0,y_0)$ and their analytical determination  \cite{prahofer2004exact}  based on the identification of the KPZ parameters below Eq.~\eqref{eq:KPZ_OG}:
\begin{align} 
    y^{\rm an}_0 = \bigg( \frac{\nu}{2\lambda^2 D}\bigg)^{1/3} & = \bigg(\frac{n}{8TJ}\bigg)^{1/3}\label{eq:y_0}\\
    C^{\rm an}_0 = \bigg(\frac{\lambda D^2}{2\nu^2}\bigg)^{2/3} & = \bigg(\frac{T^2}{n^2J}\bigg)^{2/3}\label{eq:c_0},
\end{align}
from which we obtain $y^{\rm an}_0\approx1.7$ and $C^{\rm an}_0\approx7.1\times10^{-3}$. These are far from the numerically fitted $y_0 = 8.3\times10^{-1}$ and $C_0 = 3.0\times10^{-2}$, which is not surprising given that the condition of weak density fluctuations leading to Eq. \eqref{eq:KS} is not satisfied in the grandcanonical regime.

\subsection{Crossover to canonical regime}
\begin{figure}[h!]
    \centering 
    \includegraphics[width = 1.0\linewidth]{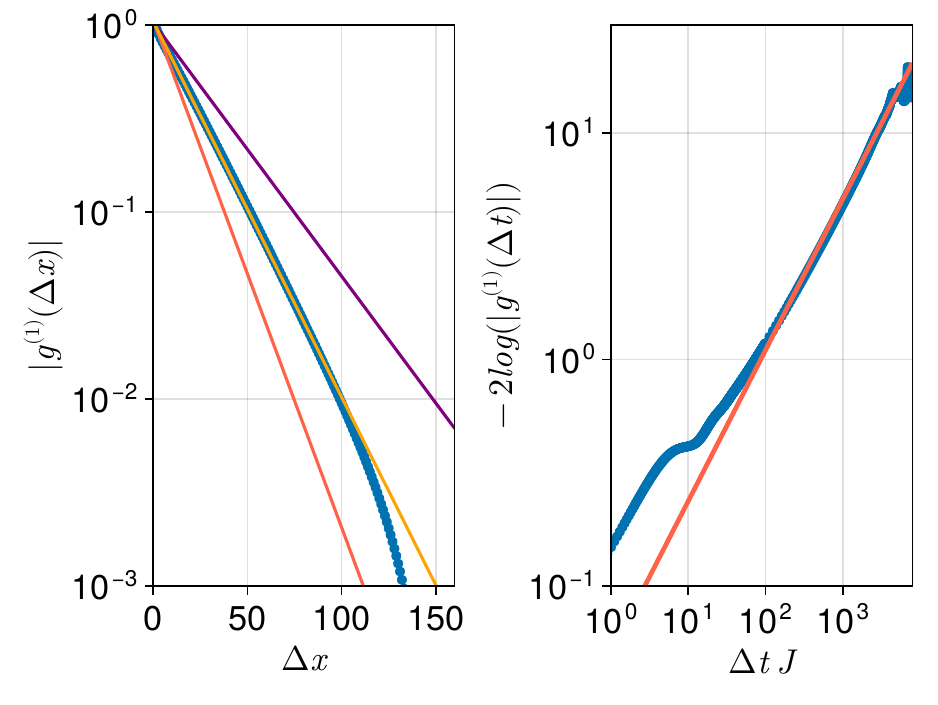}
    \caption{First order spatial (left) and temporal (right) coherence decay for a one dimensional photon condensate array in the grand canonical regime. Numeric result (blue data) from simulating eq. \eqref{eq:gGP} and \eqref{eq:rate} for equal-time spatial correlation function $g^{(1)}(x)$ (left) and equal-space temporal correlation function $g^{(1)}(t)$ (right) for a $N=1024$ site system with periodic boundaries in the canonical regime: $n^2/M_{\rm eff}\approx 1.06$. The red line depicts the NIBG exponential decay $\ell_0  = \frac{2J \bar n}{T}$ and purple line the WIBG exponential decay $2\ell_0  = 2\frac{2J \bar n}{T}$. The yellow line is obtained by fitting the exponential decay as $1.35\ell_0  = 1.35\frac{2J \bar n}{T}$. System parameters $J=0.1\:meV, \Delta=-800\:J, \gamma=\:J, B_{21}=10^{-4}\:J, n=2 \times 10^3, M=10^{8}$}
    \label{fig:spatcorcan}
\end{figure}

Let us now turn to the investigation of parameters closer to the canonical regime $n^2/M_{\rm eff}\approx 1.06$, for which we numerically obtained the quasi-condensate fraction  $n_{\rm qc}\approx 0.86$. 
With these parameters, the KPZ scaling could still be extracted with numerically accessible system size; we did not find system parameters showing KPZ scaling within our numerics deeper in the canonical regime.
The left panel of Figure \ref{fig:spatcorcan} shows the spatial coherence, which follows the expected exponential decay. The numerically determined spatial coherence however decays on a longer length scale than the NIBG coherence length $\ell_0$, but shorter than the WIBG coherence length $2\ell_c$. The intermediate value of the coherence length is in line with the simulated value of the density fluctuations $g^{(2)}(0)\approx 1.26$, between the grand canonical $g^{(2)}(0)\approx 2$ and canonical regime $g^{(2)}(0)\approx 1$.

\begin{figure}[h!]
    \centering 
    \includegraphics[width = 1.0\linewidth]{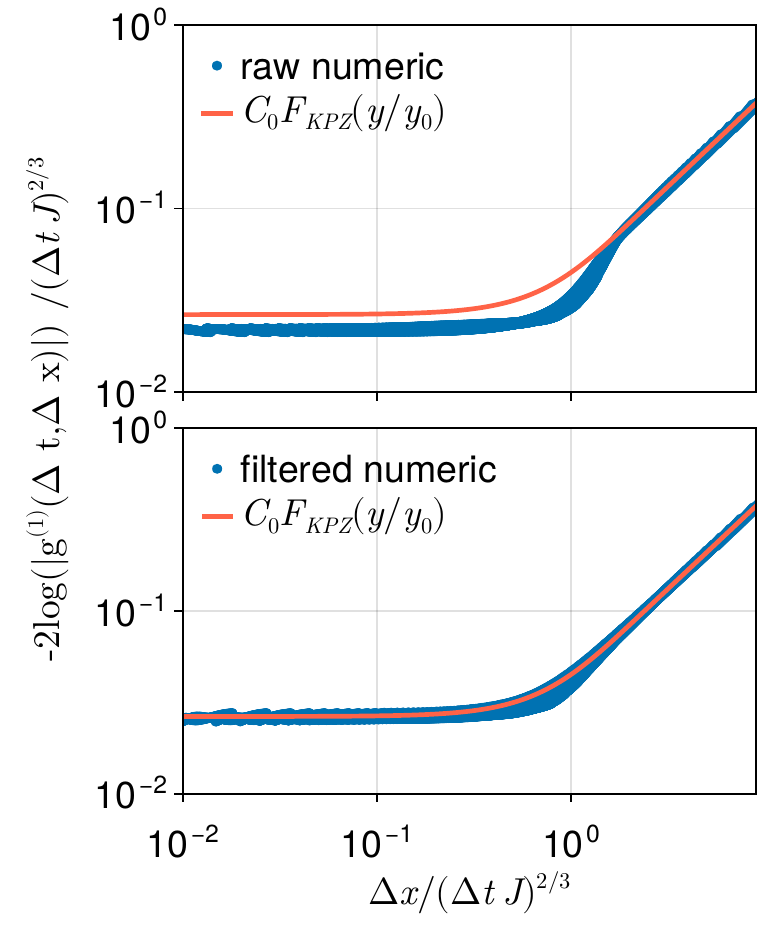}
    \caption{Correlator $C_0F_{KPZ}(y/y_0)$ collapse for spacetime points between the two contours in the left panel of Figure \ref{fig:spectrumcohmapcan}. The upper panel shows decent collapse, but not on the universal KPZ scaling function. The numerical data in the lower panel, that are obtained after applying a spectral filter shows a very good collapse on the universal KPZ scaling curve. For the filtering, we used the function described in the text, with $\omega_c=0.0016 $ and $ \delta=0.004$. Fitted parameters: $y_0\approx9.1\times10^{-1}, C_0\approx2.3\times10^{-2}$.
    Same system parameters as in Fig. \ref{fig:spatcorcan}.
    }
    \label{fig:scaling2}
\end{figure}

The scaling collapse of spacetime points in the upper panel of Figure \ref{fig:scaling2} shows a marked deviation from the universal KPZ scaling curve. Although the slope of the numerical data matches the KPZ scaling curve for small $\Delta x/(\Delta t J)^{2/3} < 0.3$ and large $\Delta x/(\Delta t J)^{2/3} > 2$, both asymptotic behaviors cannot be made to collapse onto the universal scaling curve for a single set of scaling parameters $(c_0,y_0)$.
For a value around $\Delta x/(\Delta t J)^{2/3} \approx 1$ the data points behave very differently from the universal curve.

\begin{figure}[h!]
    \begin{minipage}{.5\linewidth}
        \centering
        \includegraphics[width = 1.0\linewidth]{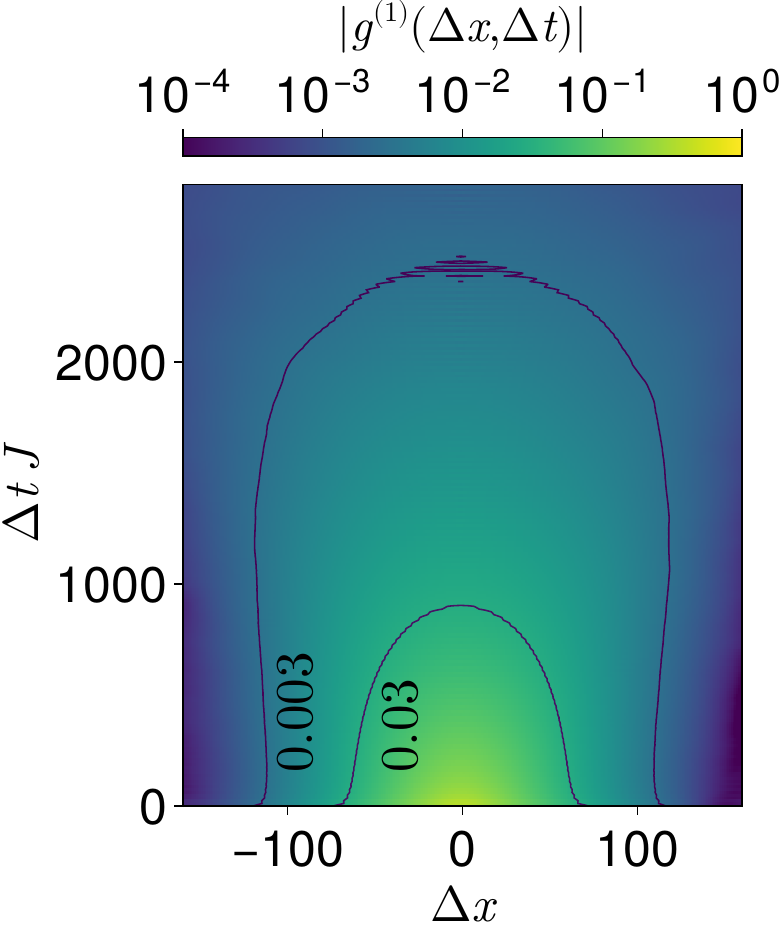}
    \end{minipage}%
    \begin{minipage}{.5\linewidth}
        \centering
        \includegraphics[width=1\linewidth]{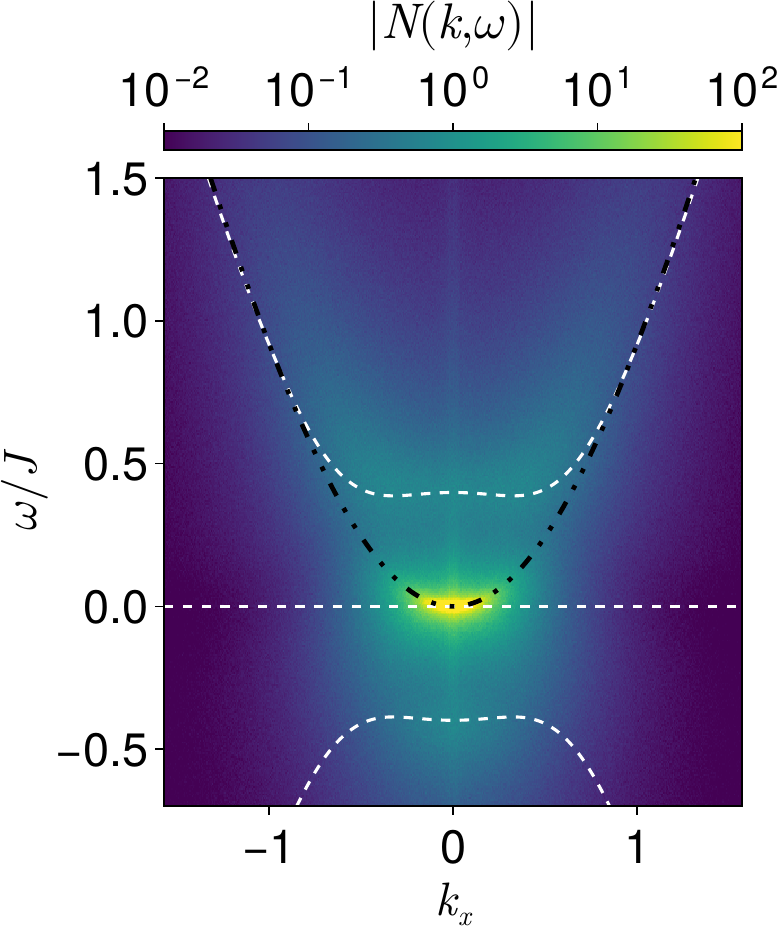}
    \end{minipage}\par\medskip
    \caption{\textit{Left panel --} Spatiotemporal first order coherence of the photon condensate in the close to canonical regime after spectral filtering as described in the text. The area inside the contours shows the spacetime points collapsing on the universal KPZ scaling curve for the canonical regime with $0.003 \le |g^{(1)}(\Delta x, \Delta t)| \le 0.03$ after filtering out high energy modes and renormalizing to $g^{(1)}(0,0)=1$ \textit{Right panel --} The energy-momentum spectrum of the photon condensate is shown with the color scale, the black dashed line represents the photon dispersion of the empty cavity, the white dashed lines represent the elementary excitation spectrum. Same system parameters as in Fig. \ref{fig:spatcorcan}.}
    \label{fig:spectrumcohmapcan}
\end{figure}

More insight in this deviation can be obtained from the  energy-momentum spectrum that is shown in Fig. \ref{fig:spectrumcohmapcan}. In contrast to the case in the grand-canonical regime (right panel of Fig. \ref{fig:spatcorgrandcan}), we see here that there is a significant contribution from states with finite momentum and energy. Part of the luminescence falls on the elementary excitation spectrum (white dashed lines) that was analytically computed with Eqs. \eqref{eq:deltan}, \eqref{eq:deltaX} and the linear part of eq.\eqref{eq:deltan}, predicting gapped particle and ghost branches as well as a flat diffusive branch.
The dominant luminescence around $k=0$ however appears to follow the quadratic single-particle TB dispersion. This feature is not captured by the linearized Bogoliubov model and indicates that nonlinear corrections are significant.

By applying a spectral filter, we can focus more on the universal long time physics. Cutting out the high energy modes of $g^{(1)}(k_x, \omega)$ by multiplying the spectrum with an exponentially decaying function $f(\omega) = \frac{1}{2}(1+\tanh{\frac{\omega - \omega_c}{\delta}})$ changes the spacetime collapse to the universal curve, as is shown in the lower panel of Fig. \ref{fig:scaling2}.  Analytical coefficients obtained from equations \eqref{eq:y_0} and \eqref{eq:c_0} in the crossover regime yields $y^{\rm an}_0\approx17$ and $C^{\rm an}_0\approx7.1\times10^{-5}$. 

These findings underscore the intrinsically diffusive character of KPZ dynamics: thanks to the fully diffusive spectrum, the dynamics manifest KPZ scaling in the grand canonical regime without additional filtering — in contrast to the behavior observed in the crossover regime.



\section{Conclusions and Outlook}\label{sec:conclusion}

In this article, we have investigated the KPZ scaling of the spatiotemporal coherence in photon condensates, that are out of equilibrium due to photonic losses, but still closely approach thermalization thanks to the repeated absorption and emission of photons by the dye molecules. For high momenta the momentum distribution shows a thermalized tail, i.e. converging to $\eta \frac{k_BT}{\epsilon_{k}}$ with $\epsilon_{k} = 2J[1-\cos{k}]$ the TB dispersion. In the long wavelength limit, where KPZ dynamics are expected, the distribution is not thermalized. 

We argued analytically within a linearized approximation for the density fluctuations that due to their nonequilibrium character, the phase fluctuations follow the scaling behavior of the KPZ universality class. This scaling behavior was confirmed numerically, even when the density fluctuations are very large.  This demonstrates the robustness of the KPZ scaling for out of equilibrium quantum fluids that undergo spontaneous $U(1)$ symmetry breaking. For intermediary density fluctuations, scaling collapse was obtained, but it markedly differed from the universal scaling curve. This discrepancy was mitigated by filtering out the high energy modes. It still remains an open question whether the scaling curve obtained without filtering also has some universal character.


Here, we used a model suited for dye filled microcavities, but more recently also photon thermalization and condensation was observed in semiconductor microcavities \cite{barland2021photon,pieczarka2024bose}. Also for these systems, KPZ scaling should be present and it will be interesting to investigate the scaling of spatiotemporal coherence in these systems.

In the current study, we have assumed a perfectly homogeneous lattice without any disorder. The effect of the disorder on the spatiotemporal scaling \cite{moroney2021synchronization} is left as a subject for further study.

\section{Acknowledgements}

We thank Martin Weitz and Julian Schmitt for stimulating discussions on photon condensation and Jacquelin Bloch, Sylvain Ravets, Quentin Fontaine, Anna Minguzzi, Léonie Canet and Félix Helluin on the KPZ physics. And a special mention to our dear colleague Vladimir N. Gladilin$^\dagger$.
MY is financially supported through the FWO-Vlaanderen through research grant G0B2923N.
\bibliography{biblio.bib}
\end{document}